\documentclass{amsart}
\usepackage{amsmath}
\usepackage{amsfonts}
\usepackage{amssymb}
\usepackage{xcolor}
\usepackage{graphicx}
\newtheorem{Theorem}{Theorem}[section]

\newcommand{\R}{\mathbb R}

\newcommand{\N}{\mathbb N}

\title[Well-posedness of generalized KP systems]{Well-posedness of generalized Kadomtsev-Petviashvili systems}

\author{Jean-Pierre Magnot${}^{*}$}

\author{Enrique G. Reyes${}^{\dagger}$}

\address{$^*$  Univ. Angers, CNRS, LAREMA, SFR MATHSTIC, F-49000 Angers, France, \\ Lepage Research Institut, 17 novembra 1, 081 16 Presov, Slovakia \\ 
	\\ and \\  Lyc\'ee Jeanne d'Arc, \\ Avenue de Grande Bretagne, \\ 63000 Clermont-Ferrand, France}
\address{$^\dagger$ Departmento de Matem\'{a}tica y Ciencia de la Computaci\'{o}n \\
	Universidad de Santiago de Chile \\ Casilla 307 Correo 2,
	Santiago, Chile.}

\email{$^*$ magnot@math.cnrs.fr ; jean-pierr.magnot@ac-clermont.fr}
\thanks{$^*$ The author thanks the France 2030 framework programme Centre Henri Lebesgue ANR-11-LABX-0020-01 
	for creating an attractive mathematical environment.}
\email{$^\dagger$ enrique.reyes@usach.cl ; e\_g\_reyes@yahoo.ca}
\thanks{$^\dagger$ Partial support from the Fondo Nacional de
Desarrollo Cient\'{\i}fico y Tecnol\'{o}gico (FONDECYT) through grant \#1241719 is gratefully acknowledged.}
\begin{document}

	\begin{abstract}
		We review our results, published, prepublished or unpublished, on the well-posedness of selected generalized Kadomtsev-Petviashvili hierarchies. 
	\end{abstract}
	
	\maketitle
	
	
\section*{Introduction}
For more than thirty-five years, the Kadomtsev-Petviashvili (KP) hierarchy and its solutions have played a central 
role in geometric approaches to Mathematical Physics, not only in the theory of integrable systems (for example 
due to its links with equations such as the Burgers, Korteweg-de Vries and Boussinesq equations, see {\em e.g.,} 
\cite{Kaz,Kodama}), but also for its connections with algebraic curves (see for instance \cite{M2}) and with the 
geometry of infinite-dimensional grassmannian manifolds, from the works of Sato to those of Pressley,Segal and 
Wilson. Considering this last issue, connecting the KP hierarchy --formulated with formal pseudodifferential 
operators-- with groups involving (non-formal) Hilbert-Schmidt operators on a separable  Hilbert space, see 
\cite{Mick,PS}, remains a exciting and amazing procedure even after all these years.	

From this last aspect of the theory of the classical KP hierarchy, the { original} KP system was 
interpreted in terms of spectral problems of infinite dimensional operators, through a tricky generalization to 
Segal-Wilson's restricted Grassmannian of Plu\"ucker coordinates on finite-dimensional Grassman manifolds. This 
correspondence has let us hope for an analytic --or at least fully differential geometric--- approach to the KP 
system first stated in a purely algebraic way on formal pseudo-differential operators, see for instance \cite{D}.

It is this journey that we summarize here, from standard settings to widely generalized systems. This 
generalization can be understood in two ways: first, by generalizing the algebra of pseudo-differential 
operators, taking formal or even non formal ones, and secondly, by generalizing the nature of the solutions, from 
solutions expressed as formal series to solutions expressed as (true) smooth functions. These investigations began 
more than eight years ago, and their results have been presented in several different publications. Here we 
provide an overview of our key findings. Perspectives are given in an outlook section.   
	
	\section{On generalized Kadomtsev-Petviashvili (KP) hierarchy on formal and non-formal pseudo-differential 
	operators}
		Let $\mathbb{K}$ be a characteristic zero diffeological field, and let  $(A,+,* ,\partial)$ be a 
		differential diffeological unital associative algebra {{} over $\mathbb{K}$.} For basics on 
		diffeological spaces, 
		see \cite{GMW2023} and references therein. 
		
		We define
	%
		$\Psi DO(A) = \left\{ \sum_{n \in \mathbb{Z}} a_n \xi^n : a_n \in A ; a_n = 0 \mbox{ for } n >> 0 
		\right\}$, in which $\xi$ is a formal variable.
	$\Psi DO(A) \subset A^\mathbb{Z}$ is a diffeological space and the classical splitting 
	$\Psi DO(A)= DO(A) \oplus IO(A)$, {{} 
	$\sum_{n \in \mathbb{Z}} a_n \xi^n \mapsto \sum_{n \geq 0} a_n \xi^n
	+ \sum_{n < -1} a_n \xi^n$,} is smooth.
		Let
		$a(\xi)=\sum a_n \xi^n$ and $b(\xi)=\sum b_n \xi^n.$ Addition and multiplication on $\Psi DO(A)$ are 
		defined thus (here and henceforth, multiplication on $A$ is denoted by concatenation rather than $*$):
		$$(a(\xi),b(\xi)) \mapsto \sum_{p \in \mathbb{Z}} (a_p+b_p)\xi^p \; \hbox{ and } \; (a(\xi),b(\xi)) \mapsto  \sum_{m,n} \sum_{\alpha\in \N} \frac{1}{\alpha !} ( a_n  \partial^\alpha b_m) (D^\alpha_\xi\xi^n)\xi^{m} 
		$$
		
	{{} These two operations }are smooth {{} (and so is inversion when defined). $\Psi DO(A)$ is 
	a diffeological unital and associative $\mathbb{K}-$algebra called the algebra of \emph{formal} 
	pseudo-differential	operators over $A$}.

	Let us now consider $Cl_{odd}(S^1,\mathbb{C}),$ the class of classical pseudo-differential operators (PDOs) of 
	integer order, see \cite{Gil}, that lie in the so-called odd-class introduced by
	 Kontsevich and Vishik \cite{KV1,KV2}.
	They are \emph{non-formal} because they are true operators acting on $C^\infty(S^1,\mathbb{C})$. 
	No such
	action exists for $\Psi DO(C^\infty(S^1,\mathbb{C}))$. 
		$Cl_{odd}(S^1,\mathbb{C})$ is a Fréchet algebra, 
		see \cite{CDMP,KV1} and \cite{Ma2022}, equipped with the topology whose open sets are generated by the 
		union of the Fr\'echet topology on formal symbols
		and the topology of off-diagonal kernels,
	and such that the formal symbol map on PDOs defines a morphism
	$Cl_{odd}(S^1,\mathbb{C}) \rightarrow \Psi DO(C^\infty(S^1,\mathbb{C})).$
	There is also a {{} natural smooth} decomposition  
	$Cl_{odd}(S^1,\mathbb{C}) = DO(S^1,\mathbb{C}) \oplus Cl^{-1}_{odd}(S^1,\mathbb{C})$, 
	see \cite{Ma2022,MR2024}.
		
		Let $T = (t_1,t_2,...)$ be a countable family of (independent) variables.
		The pseudo-differential operator $L,$ of order 1, depending on the $T-$variables, is a solution of the KP 
		hierarchy iff: 
		\begin{equation} \label{KP}
			\frac{d L}{d t_{k}} = \left[ (L^{k})_{+} , L \right] = -\left[ (L^{k})_{-} , L \right] \; , \quad
			\quad k \geq 1 \; ,
		\end{equation}
		with initial condition  $L(0) \in \partial + IO(A)$ in formal operators or 
		$L(0) \in \partial_x + Cl^{-1}_{odd}(S^1,\mathbb{C})$ in non-formal odd class operators. 
		In principle, 
		a solution to the KP hierarchy should be a smooth mapping from $\oplus_{k \in \N^*} \R t_k$ to 
		$\Psi DO^1(A)$ or to $Cl^{1}_{odd}(S^1,\mathbb{C})$.
		However, classical KP theory considers only formal solutions, {{} see \cite{D,M1,M3} or the 
		review \cite{ER}}, or germs of solutions (formal series in $T$-variables that could be understood 
		as {{} ``Taylor series" of ``true"} solutions). Therefore, when we consider solutions of the KP 
		hierarchy, without {{} further clarifications}, 
		we understand the solution of the KP hierarchy as an element of 
		{{} $\Psi DO^1(A[[T]]) \subset \Psi DO^1(A)[[T]]$ or, as we found recently,} as an element of 
		$Cl^{1}_{odd}(S^1,\mathbb{C})[[T]]$, see \cite{MR2024}.
		
		\section{Well-posedness and Zakharov-Shabat equations}
		
		\begin{Theorem} \label{KPcentral}
			Consider the KP hierarchy
			with initial condition $L(0)=L_0$ (posed on formal or non-formal operators). 
			Then, the solution $L$ (as series in $T$) exists, is unique and
			the map $L_0  \mapsto L $ is smooth.   
		\end{Theorem}
			This result has to be understood as stating the smooth dependence of the coefficients of the series $L$ 
			on the operator $L_0.$ For formal operators, the most general result of this type is, to the best of  
			our knowledge, given in \cite{ERMR} where ultrametric structures are used, in order to propose a 
			field of applications as wide as possible. It includes the classical setting in $T-$series 
			{{} considered
			in \cite{MR2016} and, in a simplified fashion,} in \cite{ERMRR}. 
	%
	%
		In classical proofs, the operator $L_0$ as well as the $T-$dependent solution $L$ are expressed as 
		$ L_0 = S_0 \partial S_0^{-1} \hbox{ and } L= S\partial S^{-1}$ for invertible operators $S_0, S$ of 
		degree 0 such that 
		$S_0 - Id$ (resp. $S-Id$) is of order $-1.$ This condition {{} was always} assumed in historical 
		groundbreaking results, e.g. in \cite{M1,M3}, or in classical textbooks such as \cite{D,Mick}. 
		It also appears in some of our theorems, see e.g. \cite{MR2025}.
		Our main strategy of proof of existence of solutions is due to Reyman and Semenov-Tian-
		Shanski \cite{RS1981} combined with Mulase's \cite{M1,M3} constructions, 
		see \cite{MR2016} and the comments appearing in \cite{ERMRR}. 
		When we consider non-formal operators, existence of solutions for formal pseudo-differential operators is 
		the first step. The control of the smoothing part is the rest of the task, performed in 
		\cite{MR2024,MR2025}.
		
		Let us highlight some selected cases: 
		\begin{itemize}
			\item $L_0 \in Cl^{1}_{odd}(S^1,\mathbb{C})$: existence of solutions is deduced from the 
			classical case of formal PDOs acting on $C^\infty(S^1,\mathbb{C})$, and the ``dressing operator" $S$ 
			 is obtained from the action of $Diff_+(S^1)$ on $Cl^{-1}_{odd}(S^1,\mathbb{C})$, see \cite{MR2024}.
			 \item Taking the formal part of these operators, we form a classical KP hierarchy (see e.g. \cite{D}) 
			 that {can be extended} to formal pseudo-differential operators with coefficients in 
			 Coulombeau distributions. 
			 According to Giordano and Wu \cite{GW2015} this is a diffeological algebra. According to a private 
			 communication by P. Giordano, the derivation operator is smooth in this diffeology and therefore the 
			 proof of \cite{ERMRR} of well-posedness of the KP hierarchy applies.
			\item $A = C^\infty(\R^3,\R^3)$ with componentwise multiplication and derivation: we can deduce a 
			Navier-Stokes-like equation from the KP hierarchy (\ref{KP}), and prove existence of solutions, see 
			\cite{MRR2022}. 
		\end{itemize} 	
	We finish with the observation that solutions 
	$L \in C^\infty(\oplus_{k \in \N^*} \R t_k , Cl^1_{odd}(S^1,\mathbb{C})) $ to (\ref{KP})
	satisfy the (zero curvature) Zakharov-Shabat (ZS) equations:
	$$
	\frac{dL^k_+}{dt_l} - \frac{dL_+^l}{dt_k} = \left[L^l_+,L^k_+\right] \quad \hbox{ and } \quad
	\frac{dL^k_-}{dt_l} - \frac{dL_-^l}{dt_k} = -\left[L^l_-,L^k_-\right].
	$$
	The second equation reads as $dZ_- - [Z_-,Z_-]=0$, which is a curvature form with values in the space $\tau_2$ 
	of Hilbert-Schmidt (HS) operators. Then, following \cite{MR2025}:
	\begin{Theorem}
		The second ZS equation is equivalent to:
		$$\forall (n,k) \in (\N^*)^2, \quad YM_{n,k}(dZ_- - [Z_-,Z_-]) = 0\, ,$$
		where $YM_{n,k}(\alpha) =  || \alpha ||_{[-n,n]^k,HS}^2\,$, the norm $|| \alpha ||_{[-n,n]^k,HS}$ is the 
		standard 
		norm on the space $\Omega^2([-n,n]^k,\tau_2)$, and 
		$[-n,n]^k= \{(t_1,...t_{2k},0,...)\, | \, \forall i, -n \leq t_i \leq n\}.$ 
	\end{Theorem}
	
	\section{Outlook}
	During the period from 2017 to now, a fast bibliographical search shows that the KP \emph{equation}, classical 
	or modified, has drawn the interest of many researchers, while the literature on the KP \emph{hierarchy} 
	during the same period is not so dense. {We believe that one of the reasons for this finding} is 
	the non-equivalence between these two approaches. 
	The KP hierarchy only produces formal solutions and currents to the KP equation, and there exist solutions of 
	the equation that cannot be reached by means of solutions of the hierarchy, see e.g. \cite{MR2016,MRR2022}. In 
	fact, the KP hierarchy seems to 
	have more applications to pure mathematical problems, see e.g. \cite{Kaz} for applications to Hodge integrals 
	and \cite{M2} for applications to the characterization of Jacobian varieties, than 
	to the problem of finding solutions to integrable systems, even if most integrable systems can be 
	derived from the KP hierarchy. 
	
	In a pioneering approach, Prykarpatski and al. \cite{PPV2024} showed that {the KP hierarchy} can 
	be helpful to solve difficult technical problems of spectral type. This 
	investigation in a \emph{non-commutative} analysis framework, suggests that the power of the KP hierarchy 
	stands on {the fact that it relies on} 
	an (arbitrarily posed) differential algebra, which is not the case of the KP equation. Therefore, {
	motivated 
	partially by \cite{MR2025}, } future 
	investigations in this direction should consider, and even mix, differential algebras and their friends on one 
	side, and (generalized) Yang-Mills like approaches on the other side.

	\end{document}